\documentclass[preprint,superscriptaddress,showpacs,preprintnumbers,nofootinbib,amsmath,amssymb,aps,floatfix]{revtex4-1}
\pdfoutput=1
\usepackage[colorlinks=true,urlcolor=cambridgeblue,linkcolor=darkraspberry,citecolor=cambridgeblue,linktocpage=true,pdfproducer=medialab,pdfa=true,anchorcolor=blue]{hyperref}

\usepackage{amsmath,amssymb,graphicx}
\usepackage{float}
\usepackage[utf8]{inputenc} 
\usepackage{xcolor}
\newcommand{\beq}{\begin{equation}}
\newcommand{\eeq}{\end{equation}}
\newcommand{\bea}{\begin{eqnarray}}
\newcommand{\eea}{\end{eqnarray}}
\newcommand{\beas}{\begin{eqnarray*}}
\newcommand{\eeas}{\end{eqnarray*}}
\newcommand{\bi}{\begin{itemize}}
\newcommand{\ei}{\end{itemize}}

\usepackage{xcolor}
\definecolor{cambridgeblue}{rgb}{0.64, 0.76, 0.68}
\definecolor{darkraspberry}{rgb}{0.53, 0.15, 0.34}

\usepackage{graphicx}
\usepackage{dcolumn}
\usepackage{bm}
\usepackage{hyperref}
\usepackage{amssymb}
\usepackage{placeins}

\usepackage{float}
\usepackage[utf8]{inputenc} 
\usepackage{xcolor}

\begin{document}
\count\footins = 1000


\title{\huge Displaced heavy neutrinos from $Z'$ decays at the LHC\ }

\author{Cheng-Wei Chiang}
\email{chengwei@phys.ntu.edu.tw}
\affiliation{Department of Physics, National Taiwan University, Taipei 10617, Taiwan}

\author{Giovanna Cottin}
\email{gfcottin@uc.cl}
\affiliation{Department of Physics, National Taiwan University, Taipei 10617, Taiwan}
\affiliation{Instituto de F\'isica, Pontificia Universidad Cat\'olica de Chile, Avenida Vicu\~{n}a Mackenna 4860, Santiago, Chile}

\author{Arindam Das}
\email{arindam.das@hetmail.phys.sci.osaka-u.ac.jp}
\affiliation{Department of Physics, Osaka University, Toyonaka, Osaka 560-0043, Japan}

\author{Sanjoy Mandal}
\email{smandal@iopb.res.in}
\affiliation{Institute of Physics, Sachivalaya Marg, Bhubaneswar 751005, India}
\affiliation{Homi Bhabha National Institute, BARC Training School Complex, Anushakti Nagar, Mumbai 400094, India}

\date{\today}

\begin{abstract}
We study the LHC sensitivity to probe a long-lived heavy neutrino $N$ in the context of $Z'$ models. We focus on displaced vertex signatures of $N$ when pair produced via a $Z'$, decaying to leptons and jets inside the inner trackers of the LHC experiments. We explore the LHC reach with current long-lived particle search strategies for either one or two displaced vertices in association with hadronic tracks or jets. We focus on two well-motivated models, namely, the minimal $U(1)_{B-L}$ scenario and its $U(1)_{X}$ extension. We find that searches for at least one displaced vertex can cover a significant portion of the parameter space, with light-heavy neutrino mixings as low as $|V_{lN}|^2\approx 10^{-17}$, and $l=e,\mu$ accessible across GeV scale heavy neutrino masses.
\end{abstract}

\maketitle

\section{Introduction} 

The observed light mass of neutrinos in the Standard Model (SM) begs for new physics explanations. The effects of such new physics are being actively looked for at the Large Hadron Collider (LHC) and several other experimental facilities worldwide. Such light neutrino masses can be explained by employing the so-called seesaw mechanism~\cite{Minkowski:1977sc,Mohapatra:1979ia,Schechter:1980gr} that in its simplest form introduces new, heavy right-handed neutrinos, which can mix with the light neutrinos in the SM. For small enough values of the mixing, the heavy neutrinos can be long-lived, leading to macroscopic decays in the LHC experiments. Such decays can be reconstructed as displaced vertices (DVs) inside the inner trackers of the LHC detectors. Growing attention to these signatures has taken place over recent years (for a recent state-of-art review of long-lived particle searches at the LHC, see Ref.~\cite{Alimena:2019zri}), as null results at the LHC may point to the possibilities that the new physics has more complex decay patterns, and that its effects may have been overlooked or misidentified by standard searches. New physics may become evident in these spectacular displaced signatures, as the SM can hardly mimic them. Their very small backgrounds will continue to make them attractive and their current study is of top importance for the high luminosity run of the LHC and future experimental facilities~\cite{Curtin:2018mvb,Ariga:2018uku,Alekhin:2015byh}.

Heavy neutrinos are predicted in several models of new physics beyond the SM. Of particular interest are the $B-L$ extensions of the SM, which have an extended gauge symmetry $U(1)_{B-L}$~\cite{Basso:2008iv} and an associated new heavy $Z'$ vector boson. The $U(1)_{B-L}$ symmetry can be broken spontaneously by the addition of a new SM-singlet Higgs field that attains a vacuum expectation value (VEV), and the theory includes three-generations of right-handed (or sterile) heavy neutrinos $N^i$, enabling the seesaw mechanism of light neutrino mass generation.  The $N^i$ can be produced from a SM-like Higgs associated with the $B-L$ breaking, or pair produced at colliders via a $Z'$. It can further decay with a displaced vertex (DV) depending on its couplings and mass. 

Dedicated experimental searches for a $Z'$ decaying to lepton pairs by the CMS collaboration placed a bound on the $Z'$ mass to be $m_{Z'}>4.5$ TeV~\cite{Sirunyan:2018exx} (assuming a SM-like gauge coupling). Recently the ATLAS collaboration analyzed the full Run 2 dataset~\cite{Aad:2019fac}, excluding a $Z'$ just below 5~TeV. For a broad review on $Z'$ models and early LHC strategies, see Ref.~\cite{Salvioni:2009mt} and references therein (see also refs.~\cite{Basso:2009hf,Basso:2010pe,Deppisch:2013cya,Das:2017deo,Jana:2018rdf,Das:2018tbd} for other collider studies in the $B-L$ case). ATLAS has now implemented a search that targets at displaced heavy neutrinos as benchmark, in the scenario where only one extra right-handed neutrino (produced in $W$ boson decays) is added to the SM~\cite{Aad:2019kiz}. Current and proposed displaced strategies in several other heavy neutrino models are an attractive focus of research in recent years~\cite{Drewes:2019fou,Boiarska:2019jcw,Cottin:2019drg,SHiP:2018xqw,Abada:2018sfh,Dercks:2018wum,Helo:2018qej,Nemevsek:2018bbt,Cottin:2018nms,Cottin:2018kmq,Dube:2017jgo,Dev:2017dui,Accomando:2017qcs,Caputo:2017pit,Caputo:2016ojx,Mitra:2016kov,Batell:2016zod,Shuve:2016muy,Antusch:2016vyf,Antusch:2016ejd,Izaguirre:2015pga,Gago:2015vma,Eijima:2018qke,Ibarra:2018xdl}.

LHC constraints for the minimal $U(1)_{B-L}$ model were addressed through a global fit in Ref.~\cite{Amrith:2018yfb} for several choices of model parameters, but signatures involving displaced heavy neutrinos were not considered.  Recent works on displaced neutrinos in $U(1)_{B-L}$ models have focused on displaced signatures coming from Higgs bosons due to a higher production cross section~\cite{Accomando:2016rpc,Deppisch:2018eth}. For this reason, production via a $Z'$ has had less attention. Early displaced strategies for a simplified model were recast in Ref.~\cite{Batell:2016zod}, with focus on a benchmark scenario with relatively unboosted $N$. Recently, the authors in Ref.~\cite{Deppisch:2019kvs} estimated the reach of future lifetime frontier experiments (like FASER \cite{Kling:2018wct, Ariga:2018uku} and MATHUSLA \cite{Curtin:2018mvb}) on a rather light $N$ and $Z'$, of $\mathcal{O}$(GeV) masses. In this work, we focus on higher masses and the LHC capabilities running at high luminosity, by reinterpreting ongoing DV searches at ATLAS and CMS. We also investigate prospects in a more general scenario than the $B-L$ model, the so-called non-exotic $U(1)_{X}$ extension of the SM~\cite{Appelquist:2002mw}, as it has been shown in Refs.~\cite{Das:2017flq,Das:2017deo} that an enhancement in the $Z'$ production is possible, providing increased sensitivity to more complex scenarios in the search for displaced heavy neutrinos when they come from a $Z'$. The lifetime of $N$ as a function of the lightest neutrino mass under the general $U(1)_X$-extended models have been studied in Refs.~\cite{Das:2018tbd,Das:2019fee}.

The paper is organized as follows. We summarize the model under study in the Sec.~\ref{sec:model}.  We also extract the exclusion region in the parameter space of new gauge coupling and $Z'$ mass using the Drell-Yan processes measured by both ATLAS and CMS. In Sec.~\ref{sec:LHCDV}, we discuss the ATLAS and CMS displaced searches, reinterpret their results, and identify discovery prospects at the high luminosity LHC. We summarize and conclude in Sec.~\ref{close}.

\section{The Model \label{sec:model}}

We consider an extension of the SM to have the gauge group $SU(3)_c \times SU(2)_L \times U(1)_Y \times U(1)_X$, where $U(1)_X$ is realized as a linear combination of the SM $U(1)_Y$ and the $U(1)_{B-L}$ symmetries~\cite{Mohapatra:1980qe}, known as the non-exotic $U(1)_X$ extension of the SM~\cite{Appelquist:2002mw}. The model is free from all the gauge and mixed gauge-gravity anomalies due to the presence of three generations of right-hand neutrinos (RHNs) $N^i$ (with $i=1,2,3$)~\cite{Das:2016zue,Oda:2015gna}. A new scalar field $\Phi$ is also introduced to break the $U(1)_X$ symmetry by attaining a VEV. The particle content is given in table~\ref{pc}. 
\begin{table}[h]
\begin{center}
\begin{tabular}{c|ccc|r}
\hline
            & SU(3)$_c$ & SU(2)$_L$ & U(1)$_Y$ & {$U(1)_X$} \\
\hline
$q_L^i$    & {\bf 3}   & {\bf 2}& $+1/6$ &  $\frac{1}{6}x_H + \frac{1}{3}x_\Phi$  \\[2pt] 
$u_R^i$    & {\bf 3}   & {\bf 1}& $+2/3$ &  $\frac{2}{3}x_H + \frac{1}{3}x_\Phi$  \\[2pt] 
$d_R^i$    & {\bf 3}   & {\bf 1}& $-1/3$ &  $-\frac{1}{3}x_H + \frac{1}{3}x_\Phi$  \\[2pt] 
\hline
$\ell_L^i$& {\bf 1} & {\bf 2}& $-1/2$ &  $- \frac{1}{2}x_H - x_\Phi$   \\[2pt] 
$e_R^i$   & {\bf 1} & {\bf 1}& $-1$   & $- x_H - x_\Phi$  \\[2pt] 
\hline
$H$       & {\bf 1} & {\bf 2}& $+1/2$ & $\frac{1}{2} x_H$ \\ 
\hline
$N^i$   & {\bf 1} & {\bf 1}& $0$    & $- x_\Phi$ \\[2pt] 
$\Phi$    & {\bf 1} & {\bf 1}& $0$    & $2x_\Phi$  \\ 
\hline
\end{tabular}

\end{center}
\caption{
The particle content and gauge charges of the non-exotic $U(1)_X$ model~\cite{Appelquist:2002mw}.  The choice of $x_H=0$ and $x_\Phi=1$ corresponds to the $B-L$ case.
}
\label{pc}
\end{table}

The charges of the particles are controlled by two parameters only, $x_H$ and $x_\Phi$, as seen in table~\ref{pc}. As the $U(1)_X$ gauge group can be defined as a linear combination of the SM $U(1)_Y$ and the $U(1)_{B-L}$, by setting $x_H=0$ and $x_\Phi=1$, we recover the minimal $B-L$ scenario~\cite{Davidson:1978pm,Mohapatra:1980qe}. Therefore, without loss of generality, we fix $x_\Phi=1$ in our analysis throughout the paper. 

The Yukawa sector of the model can be written in a gauge-invariant way as 
\begin{eqnarray}
{\cal L}_{Y} &=& - \sum_{\alpha, \beta=1}^3 Y_u^{\alpha \beta} \overline{q_L^\alpha} \tilde{H} u_R^\beta
                                - \sum_{\alpha, \beta=1}^3 Y_d^{\alpha \beta} \overline{q_L^\alpha} H d_R^\beta 
                                - \sum_{\alpha, \beta=1}^3 Y_e^{\alpha \beta} \overline{\ell_L^\alpha} H e_R^\beta \nonumber \\                                
			&&- \sum_{\alpha, \beta=1}^3 Y_D^{\alpha \beta} \overline{\ell_L^\alpha} \tilde{H} N^\beta 
			- \sum_{\alpha=1}^3 Y_N^\alpha \Phi \overline{N^{\alpha C}} N^\alpha + {\rm h.c.},
\label{LYk}
\end{eqnarray}
where $\alpha,\beta$ are generation indices, $\tilde{H} \equiv i  \tau^2 H^*$, and $C$ denotes the charge conjugation. The fourth and fifth terms in eq.~\eqref{LYk} are the Dirac and Majorana Yukawa terms. Without loss of generality, we use a diagonal basis for the Majorana Yukawa matrix. After the breaking of the $U(1)_X$ and the electroweak symmetries, the $U(1)_X$ gauge boson $Z^\prime$ mass, the Majorana and neutrino Dirac masses are generated, given by
\bea
 M_{Z^\prime} &=& g_1^\prime \sqrt{4 v_\Phi^2+  \frac{1}{4}x_H^2 v_{\rm SM}^2} \simeq 2 g_1^\prime v_\Phi , \nonumber \\ \; \;
 M_{N^\alpha}&=&\frac{Y_{N}^\alpha}{\sqrt{2}} v_\Phi, \nonumber \\ \; \; 
 M_{D}^{\alpha \beta}&=&\frac{Y_{D}^{\alpha \beta}}{\sqrt{2}} v_{\rm SM},
\label{masses}   
\eea   
where $g_1^\prime$ is the $U(1)_X$ gauge coupling, $v_\Phi$ is the VEV of $\Phi$, and $v_{\rm SM}=246$~GeV is the SM Higgs VEV. To be in agreement with LEP constraints, we take $v_\Phi \gg v_{\rm SM}$~\cite{Carena:2004xs, Heeck:2014zfa}. LHC constraints will be discussed shortly.

In this model, through the $U(1)_X$ symmetry breaking, the Majorana mass terms for the RHNs are generated, and the light neutrinos acquire masses via the seesaw mechanism~\cite{Minkowski:1977sc,Schechter:1980gr}. The neutrino mass matrix is given by
\bea
m_\nu=
\begin{pmatrix}
0&M_D\\
M_D^T&M_N
\end{pmatrix}.
\label{mass-mat}
\eea
Considering $|M_D^{\alpha \beta}/M_N^{\alpha}| \ll 1$ and diagonalizing the neutrino mass matrix in eq.~\eqref{mass-mat}, we obtain the light neutrino mass eigenvalue matrix as
\bea
m_\nu\simeq - M_D M_N^{-1} M_D^T.
\label{eq:lightnu}
\eea

The heavy neutrinos are SM gauge singlets. They interact with the $W$ and $Z$ bosons via mixing with the SM neutrinos. As a result, the SM neutrino flavor eigenstate $(\nu)$ can be expressed as a linear combination of the light mass eigenstates $(\nu^m)$ and heavy mass eigenstates $(N^n)$:

\bea 
\nu_\ell \simeq  U_{\ell m}\nu^m  + V_{\ell n} N^n,  
\eea 
where $\ell$ and $m$ are generation indices, $U_{\ell m}$ is the $3\times 3$ light neutrino mixing matrix and it is identical to the PMNS matrix at the leading order (ignoring effects of non-unitarity), and 
\bea
V_{\ell n} \simeq M_D M_N^{-1}
\label{eq:mixing}
\eea
is the mixing between the SM neutrinos and the heavy neutrinos assumed to be much less than 1. 

The charged-current interactions can be expressed in terms of the  neutrino mass eigenstates as 
\bea 
{\mathcal{L}_{CC} \supset 
 -\frac{g}{\sqrt{2}} W_{\mu}
  \bar{e} \gamma^{\mu} P_L   V_{\ell n} N^n  + \rm{h.c}.}, 
\label{CC}
\eea
where $e$ represents the three generations of the charged leptons, and $P_L =\frac{1}{2} (1- \gamma_5)$ is the projection operator. Similarly, in terms of the mass eigenstates, the neutral-current interactions are written as
\bea 
{\mathcal{L}_{NC} \supset 
 -\frac{g}{2 c_w}  Z_{\mu} 
\left[ 
  \overline{N}^m \gamma^{\mu} P_L  (V^{\dagger} V)_{mn} N^n 
+ \left\{ 
  \overline{\nu}^m \gamma^{\mu} P_L (U^{\dagger}V)_{mn}  N^n
  + \rm{h.c.} \right\} 
\right] , }
\label{NC}
\eea
 where $c_w \equiv \cos \theta_w$ with $\theta_w$ being the weak mixing angle.

Due to the nonzero $U(1)_X$ charges, the $Z^\prime$ boson interacts with the particles in the same way as it does in the $B-L$ scenario. However, the couplings of such interactions will depend upon the $x_H$ and $x_\Phi$ parameters. As we have already fixed $x_\Phi=1$, the corresponding partial decay widths of $Z^\prime$ into fermions, which are of interest in this work, will depend upon the choice of $x_H$. The expressions are given in the Appendix.

The interaction between the $Z^\prime$ and the quarks can be written as
\bea
\mathcal{L}_{int} = -g'_1 (\overline{q_L}\gamma_\mu Q_{x_{L}}^q q_L+ \overline{q_R}\gamma_\mu Q_{x_{R}}^q  q_R) Z_\mu^\prime,
\eea
where $q_L~(q_R)$ is the left- (right-) handed quark and $Q_{x_{L}}^q~(Q_{x_{R}}^q)$ is the corresponding $U(1)_X$ charge. 
The corresponding interaction between the lepton sector and $Z^\prime$ can be written as
\bea
\mathcal{L}_{int} =-g'_1 (\overline{\ell_L}\gamma_\mu Q_{x_{L}}^\ell  \ell_L+ \overline{e_R}\gamma_\mu Q_{x_{R}}^\ell  e_R)Z_\mu^\prime,
\eea 
where $\ell_L~(e_R)$ is the left- (right-) handed lepton and the $Q_{x_{L}}^\ell (Q_{x_{R}}^\ell)$ is the corresponding $U(1)_X$ charge. All these charges are given in Table~\ref{pc} as a function of $x_H$ and $x_\Phi$. 

For the rest of the paper, we choose two values for $x_H$: (a) $x_H=0$, that corresponds to the $B-L$ case, or (b) $x_H=-1.2$, which is the value found to maximize the branching ratio of $Z'$ to RHNs~\cite{Das:2017flq}. These charge assignments determine our labels for the $U(1)_{B-L}$ and $U(1)_{X}$ models in what follows. We also set the mixing between the SM Higgs and the new scalar $\Phi$ to zero, as this ensures pair production of the RHNs through a $Z'$.

With the above choices, we implement both the $U(1)_X$ and $U(1)_{B-L}$ models in the {\textsc{UFO}} format~\cite{Degrande:2011ua}. We first study the LHC process $pp \to Z^\prime \to \ell^+ \ell^-$ for $\ell= e, \mu$. We reinterpret the latest ATLAS~\cite{Aad:2019fac} and CMS~\cite{Sirunyan:2018exx} limits on both models, after validating the exclusions for the sequential $Z'_{SSM}$ model the experiments present. For CMS, we consider the last 13-TeV result available, corresponding to an integrated luminosity of 36/fb. For ATLAS we make use of the latest full Run-II result with 139/fb. 

\begin{figure}[h]
\centering
\includegraphics[width=0.7\textwidth,angle=0]{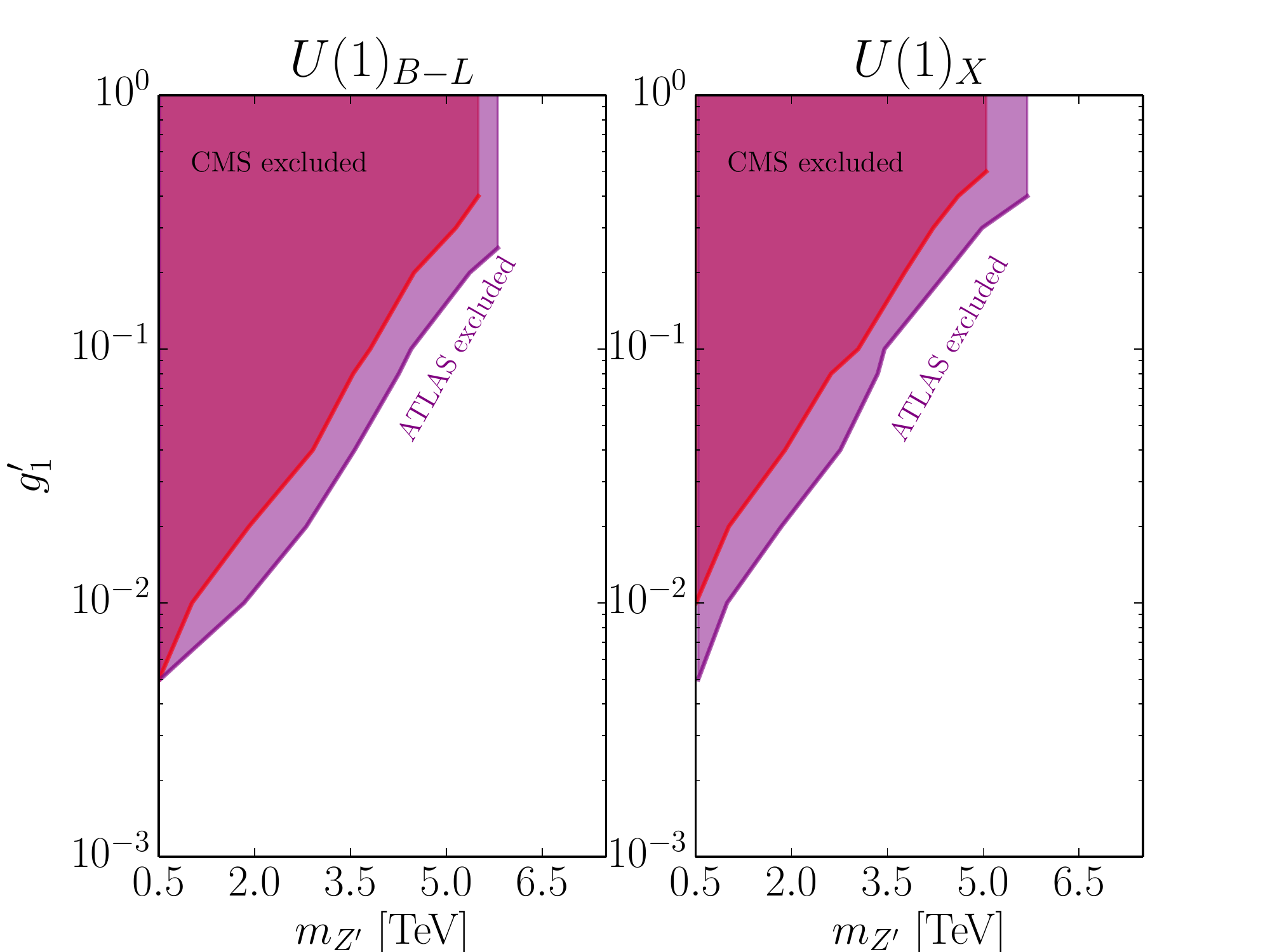}
\caption{Reinterpreted $95\%$ CL exclusion limits based on the ATLAS~\cite{Aad:2019fac} and CMS~\cite{Sirunyan:2018exx} results in the search for a heavy resonance decaying to lepton pairs, in the context of the $U(1)_{B-L}$ (left) and $U(1)_{X}$ (right) models.}
\label{fig:dilepton}
\end{figure}

Events are generated for different $(m_{Z'},g'_{1})$ values for both models with \textsc{MadGraph5}~\cite{Alwall:2014hca} at the leading order. By comparing with the combined experimental upper limits (Figure 3b of Ref.~\cite{Aad:2019fac} at $\Gamma/m = 0.5\%$, and Figure 3c of Ref.~\cite{Sirunyan:2018exx}), we can obtain lower (upper) bounds on $m_{Z'}$ ($g'_{1}$). Figure~\ref{fig:dilepton} shows the excluded region at $95\%$ CL. Our ATLAS result is consistent with Figure 3a of Ref.~\cite{Amrith:2018yfb}, performed with their previous dataset.

We use these bounds to fix a benchmark scenario of $(m_{Z'},g'_{1})= (6$ TeV, $0.8$) for the rest of the paper, consistent with current resonance searches at the LHC.

We now focus on heavy neutrino production and decay at the LHC via a $Z'$. We scan $m_{N}$ from 20 GeV to 1200 GeV, and $|V_{lN}|^2$ from $10^{-5}$ to $10^{-20}$. Existing constraints in the light-heavy neutrino mixing and mass plane for $m_{N}\geq 20$ GeV exclude $|V_{lN}|^2>10^{-6}$. LEP data excludes $|V_{lN}|^2>10^{-5}$~\cite{Deppisch:2015qwa}, where DELPHI places limits on right-handed states produced in $Z$ boson decays. Prompt collider LHC searches for three leptons in the final state exclude mixings $|V_{\mu N}|^2>10^{-5}$~\cite{Sirunyan:2018mtv,Aad:2019kiz}. Also, the ATLAS displaced search for sterile neutrinos places bounds on $|V_{\mu N}|^2$ to be less than $10^{-6}$~\cite{Aad:2019kiz} for masses of tens of GeV. Neutrinoless double beta decay ($0\nu\beta\beta$) experiments such as GERDA can also constrain $|V_{e N}|^2$. In Ref.~\cite{Cottin:2018nms}, an updated limit in the case where the SM is extended with only one right-handed neutrino produced in the decay of $W$ boson excludes values above $10^{-7}$. We will show in the next section that strategies presented in this work are more stringent than all of these constraints.

\begin{figure}[h]
\centering
\includegraphics[width=0.6\textwidth,angle=0]{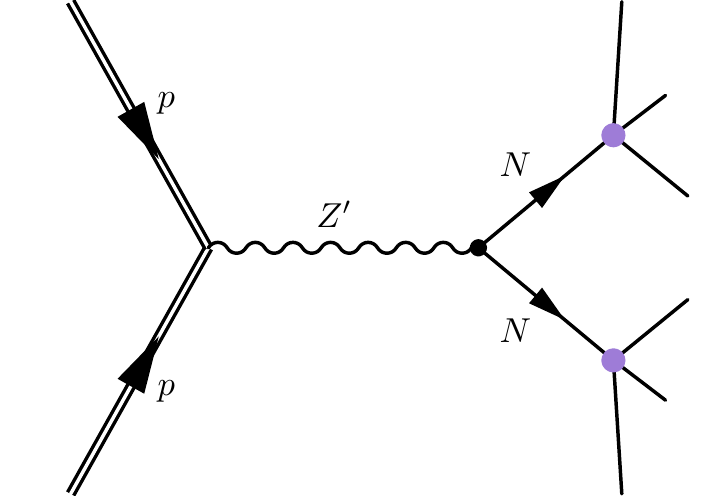}
\caption{ Heavy neutrino $N$ production and decay at the LHC in the $Z'$ models considered in this work. Production proceeds through a $Z'$, which can be either on-shell or off-shell. The neutrino then decays to a lepton (either $e$ or $\mu$) and jets, giving visible charged decay products coming from the DV. The DV positions are marked by purple circles.}
\label{FeynmanBL}
\end{figure}

\begin{figure}[h]
\centering
\includegraphics[width=0.7\textwidth,angle=0]{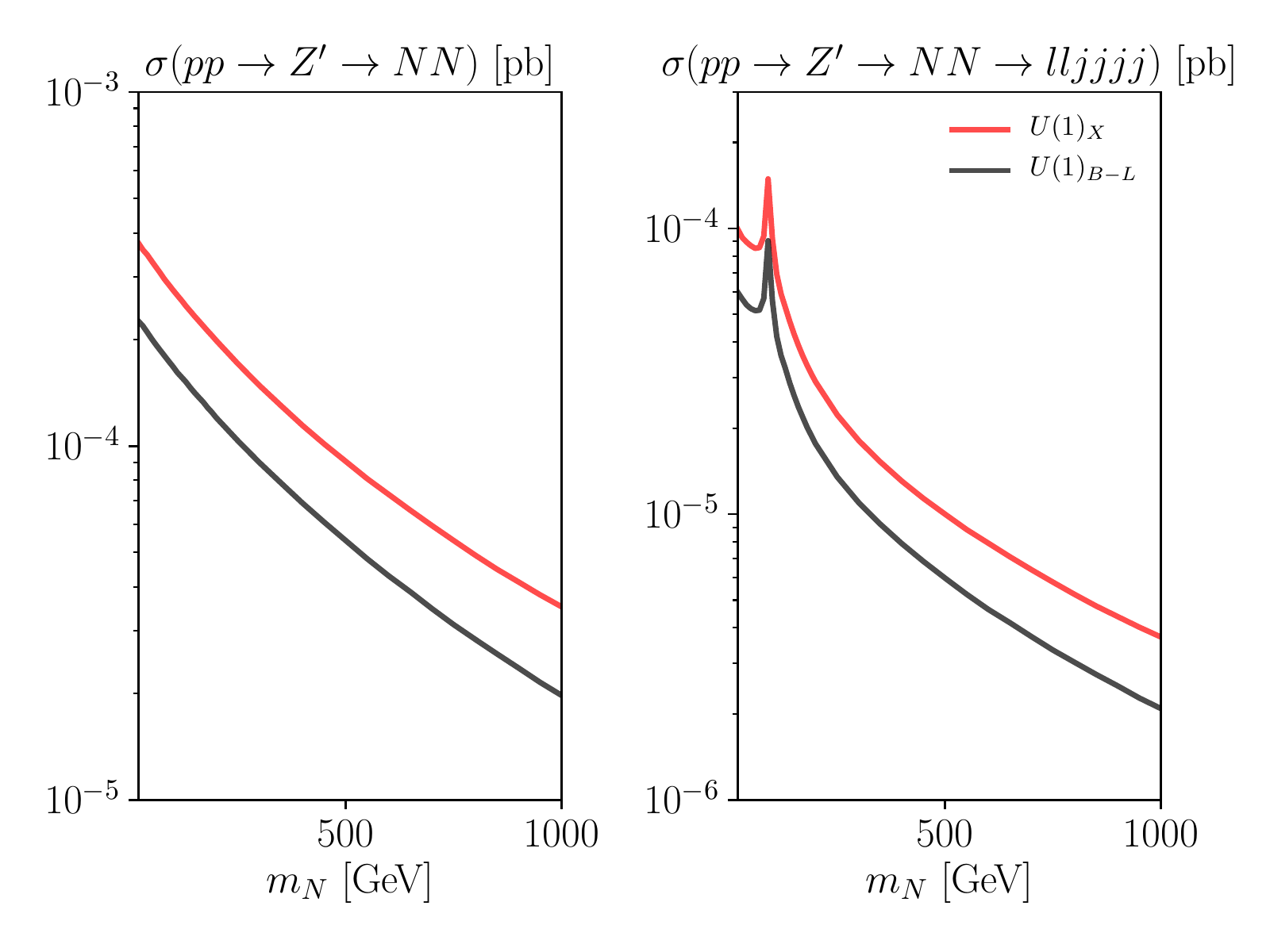}
\caption{ Pair production cross section of right handed neutrinos (left) and that including their further decays to leptons and jets (right) as a function of the heavy neutrino mass $m_{N}$, for $(m_{Z'},g'_{1})=(6$ TeV, $0.8)$. Curves for both the $U(1)_X$ and $U(1)_{B-L}$ models are shown.}
\label{fig:sigma}
\end{figure}

A diagram showing the $N$ production and displaced decay considered here is presented in Figure~\ref{FeynmanBL}. The cross section of heavy neutrino pair production and that with heavy neutrino decays are shown in Figure~\ref{fig:sigma} for both models.


\section{LHC sensitivity with displaced vertex searches}
\label{sec:LHCDV}

We focus on light-heavy neutrino mixing in the electron and muon sector separately ($V_{lN}$, with $l=e, \mu$); so only one $N^i$ will couple to either electrons or muons. In both cases, only one neutrino, now defined as $N^i=N$, is in the kinematical region of interest. We adapt the \textsc{UFOs} so that the light-heavy neutrino mixing ($V_{lN}$) and the sterile neutrino masses ($m_{N}$) are treated as independent parameters. 

We study the production process $pp\to Z'\to N N$, with each heavy neutrino decaying via $N\to l jj$ and $l=e$ or $\mu$. Event generation is performed with \textsc{MadGraph5}~\cite{Alwall:2014hca} at the leading order and the decay of $N$ is processed with {\textsc{MadSpin}}~\cite{Artoisenet:2012st}. We save the decay information by setting the {\texttt{time\_of\_flight}} variable inside \textsc{MadGraph5}'s $\texttt{run\_card}$. The generated events are then interfaced to \textsc{Pythia8 v2.3}~\cite{Sjostrand:2014zea} for showering and hadronization. Plots are generated with {\texttt{matplotlib}}~\cite{Hunter:2007}.

\begin{figure}[h]
\centering
\includegraphics[width=0.8\textwidth,angle=0]{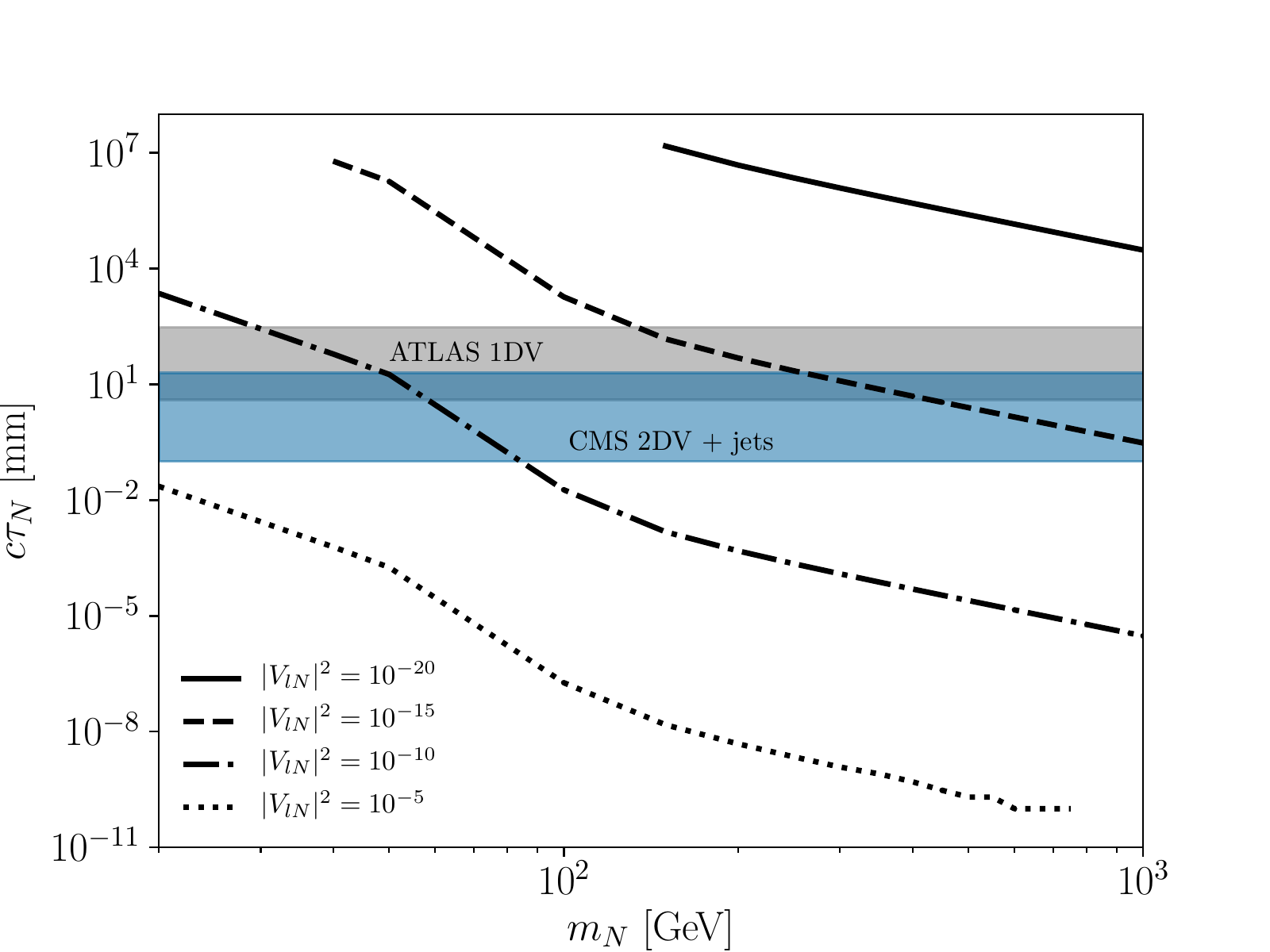}
\caption{Proper decay distance $c\tau_{N}$ as a function of $N$ mass and light-heavy neutrino mixing $|V_{lN}|^2$. The shaded region represents roughly the region that can be accessed with current DV searches at CMS~\cite{Sirunyan:2018pwn} and the ATLAS inner tracker~\cite{Aaboud:2017iio}.}%
\label{fig:ctau}
\end{figure}

We first reconstruct objects at the truth level and then apply efficiency corrections depending on the displaced search strategy of interest. We consider two complementary searches, each sensitive to a particular proper lifetime of the heavy neutrino. In Figure~\ref{fig:ctau} we show the naive reach in the proper lifetime and mass plane, for fixed mixing. We label the searches that we recast as: the ‘‘ATLAS 1DV ID" search for at least one DV in the ATLAS Inner Detector~\cite{Aad:2015rba} and the ‘‘CMS 2DV+jets"~\cite{Sirunyan:2018pwn} search for DV pairs in association with jets.
The former is a proposal inspired by searches for multitrack DVs at ATLAS~\cite{Aad:2015rba,Aaboud:2017iio}. The latter is a direct reinterpretation of the CMS search for two DVs in the inner tracker and jets~\cite{Sirunyan:2018pwn}. 

\subsection{Reach with ATLAS 1DV ID}

The 8-TeV ATLAS 1DV ID search~\cite{Aad:2015rba} looks for high track-multiplicity DVs in events possessing at least one DV and either leptons, jets or missing transverse momenta. The 13-TeV version of the search in Ref.~\cite{Aaboud:2017iio} only triggers on missing transverse momenta, but provides a prescription to implement parametrized efficiencies for DVs as a function of the vertex invariant mass and number of charged tracks. Therefore, in this work we propose a 13-TeV search with the same displaced reconstruction but with a lepton trigger (as in the 8-TeV search). 

The DV reconstruction in both 8- and 13-TeV searches is very similar to the experiment's recent search for displaced heavy neutrinos in Ref.~\cite{Aad:2019kiz}. Tunes of the former ATLAS multitrack search to target at displaced neutrinos were proposed in Ref.~\cite{Cottin:2018kmq}, matching closely the actual experimental cuts in Ref.~\cite{Aad:2019kiz}. Our vertex implementation in this work is basically the same as done in Ref.~\cite{Cottin:2018kmq}. 

The DV is reconstructed inside the ATLAS inner tracker from all charged particle tracks. The search triggers on a lepton, which is required to be associated with the DV. Cuts on the invariant mass of the vertex $m_{{DV}}$ and the amount of charged particle tracks $N_{{trk}}$ coming from it ensure this search to be free of backgrounds\footnote{Backgrounds to this search are mostly instrumental, from random crossings of tracks that fake a DV, and the background from heavy flavour, which are mostly $B$ hadrons that merge to fake vertices, or are crossed by a random track, making a high mass ``fake" vertex.}. We follow the same optimized cuts for $m_{{DV}}$ and $N_{ {trk}}$ justified in Ref.~\cite{Cottin:2018kmq}.

We have to identify electrons, muons, tracks and DVs. For muons (electrons), we use a flat identification efficiency of $90\%$ ($70\%$). We apply vertex-level efficiencies as a function of DV distance, mass and number of tracks, using the parametrized selection efficiencies from the ATLAS 13-TeV DV search~\cite{Aaboud:2017iio}. A validation study for these efficiencies was performed for the Les Houches PhysTeV 2017 proceedings in Ref.~\cite{Brooijmans:2018xbu}. 

Events are first selected by triggering on a lepton. For electrons we require $p_{T}>120$~GeV within $|\eta|<2.47$. For muons, we demand $p_{T}>55$~GeV and within $|\eta|<1.07$. DVs are selected by reconstructing tracks with a large transverse impact parameter\footnote{The transverse impact parameter is defined as $d_{0}=r\times\Delta\phi$, with $r$ being the transverse distance of the track and $\Delta\phi$ being the azimuthal angle between the decay product and the trajectory of the long-lived neutrino.}, $|d_{0}|>2$ mm and with $p_{T}>1$ GeV. The vertex position must be between 4~mm and 300~mm  and must have at least 4 tracks, $N_{ {trk}}\geq 4$. The lepton that fires the trigger must be associated with the DV. We account for this by truth-matching the lepton index with one of the displaced tracks. Note that no isolation requirement is applied on leptons. Finally, the invariant mass of the DV, $m_{ {DV}}$, is calculated assuming all tracks have the mass of the pion and it is required to be larger than 5~GeV. A summary of all selections can be found in Table~\ref{tab:selections}.

\begin{table}[ht]
\begin{center}
\begin{tabular}{p{3.0cm}|p{11.5cm}}
\hline
Trigger          &  Muon:  $|\eta| <1.07$ and $p_{T} > 55$ GeV \\
                 & Electron: $|\eta|<2.47$ and $p_{T} > 120$ GeV \\
DV region        & DV within $4$ mm $<r_{ {DV}}<300$ mm and $|z_{ {DV}}|<300$ mm\\
DV selection     & Made from tracks with $|d_{0}|>2$ mm and with $p_{T}>1$ GeV 

DV track multiplicity $N_{ {trk}}\geq 4$ and invariant mass $m_{ {DV}} \geq 5$ GeV\\
\hline
\end{tabular}
\end{center}
\caption{\label{tab:selections} Cuts for the ATLAS 1DV ID proposed search. These are optimized as in Ref.~\cite{Cottin:2018nms}, and are inspired by the ATLAS search~\cite{Aad:2015rba}.}
\end{table}

\begin{figure}[h]
\centering
\includegraphics[width=0.7\textwidth,angle=0]{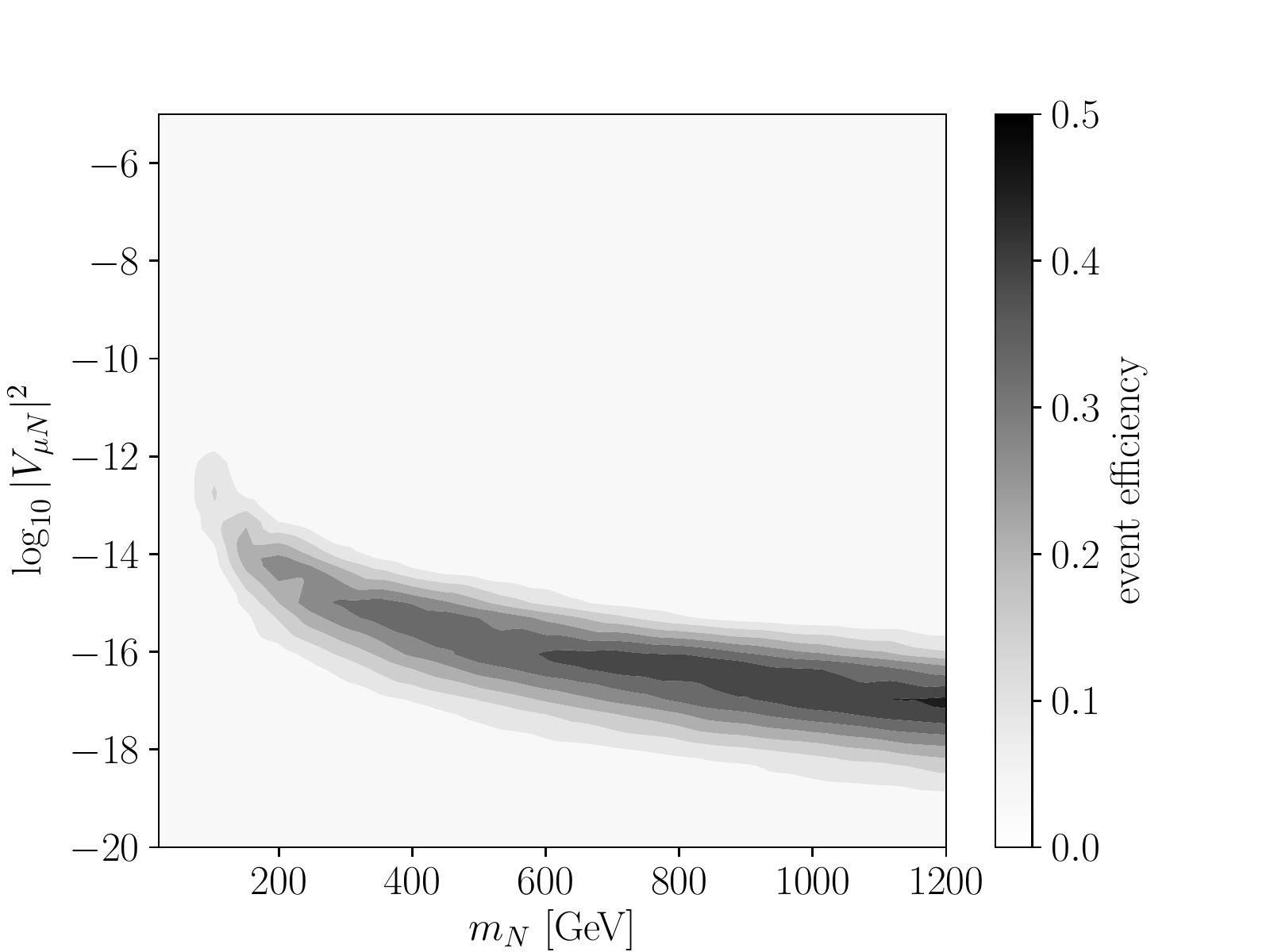}
\caption{Representative event level efficiency of the ATLAS 1DV ID search as a function of $m_{N}$ and $|V_{\mu N}|^{2}$. }
\label{fig:ATLAS1DVeff}
\end{figure}

The event level efficiency of this strategy, after all selections, is shown in Figure \ref{fig:ATLAS1DVeff} for the $U(1)_{X}$ model. The efficiency has a cigar-like shape, and is bounded by the case when the neutrinos are decaying either too promptly or too far away, outside of the detector's acceptance. For a fixed mass, such as $m_{N}=100$~GeV, and mixings bigger than $\sim 10^{-10}$, the neutrino already decays too promptly. The efficiency in this case goes down with increasing mixing for a fixed heavy neutrino mass. For fixed mixing and smaller masses, the neutrinos are decaying outside of the tracker's acceptance.

\begin{figure}[h]
\centering
\includegraphics[width=0.45\textwidth,angle=0]{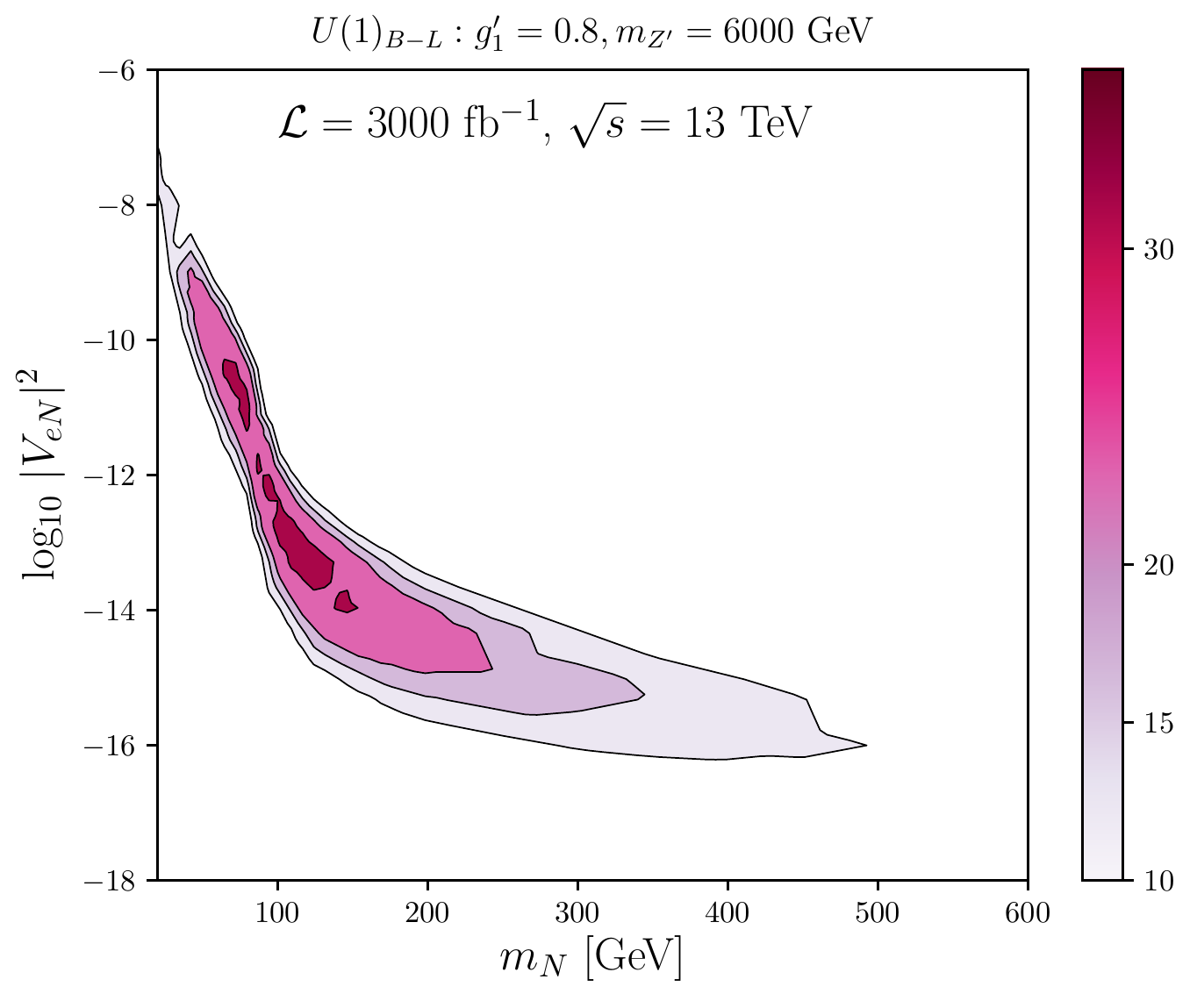}
\includegraphics[width=0.45\textwidth,angle=0]{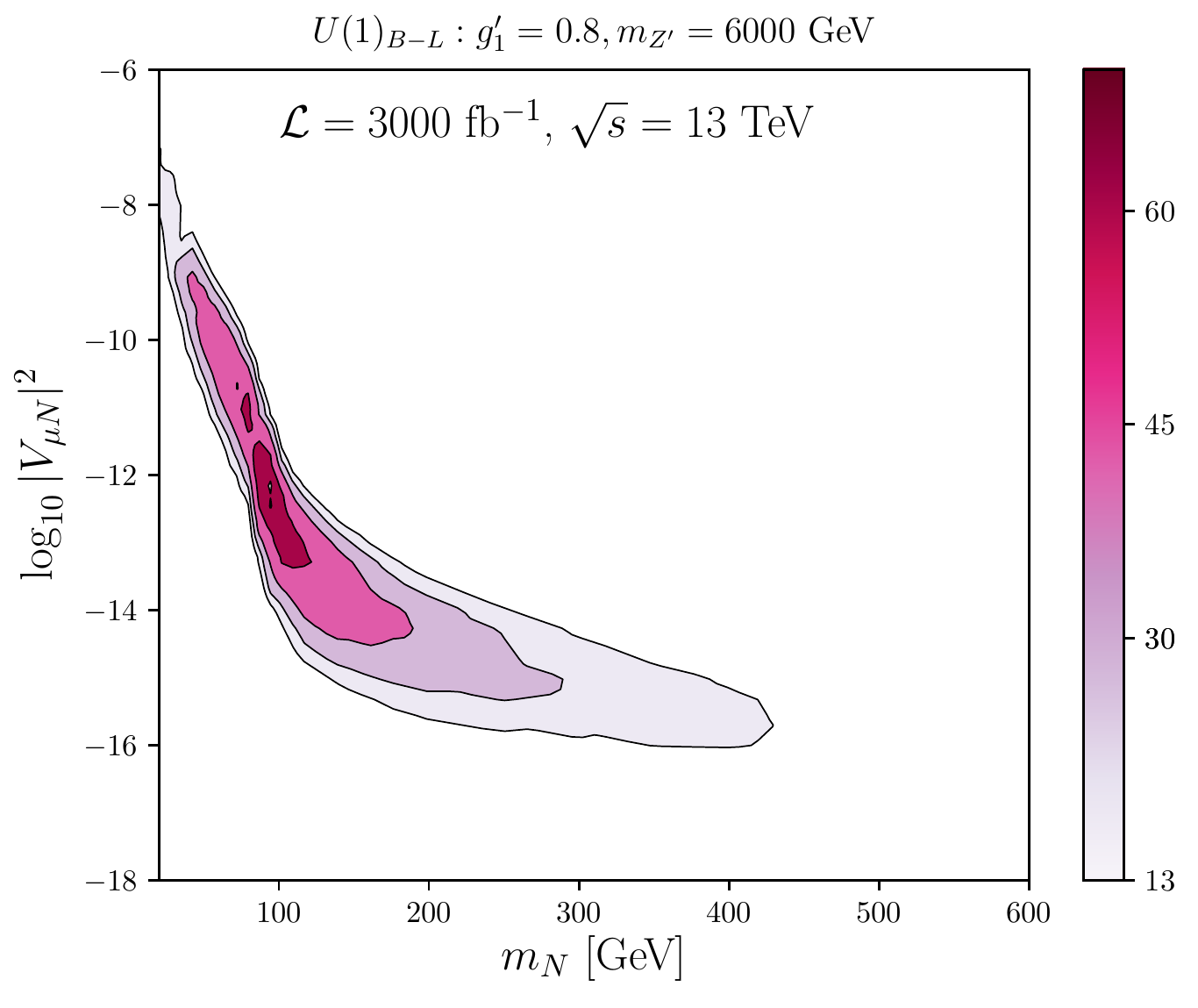}
\caption{Number of signal events for the $U(1)_{B-L}$ model at $\sqrt{s}=13$ TeV expected in $\mathcal{L}=3000$ fb$^{-1}$ with the ATLAS 1DV ID search. Reach for mixings in the electron (left) and muon (right) sector are shown. }
\label{fig:nEventsATLAS1DVBL}
\end{figure}

\begin{figure}[h]
\centering
\includegraphics[width=0.45\textwidth,angle=0]{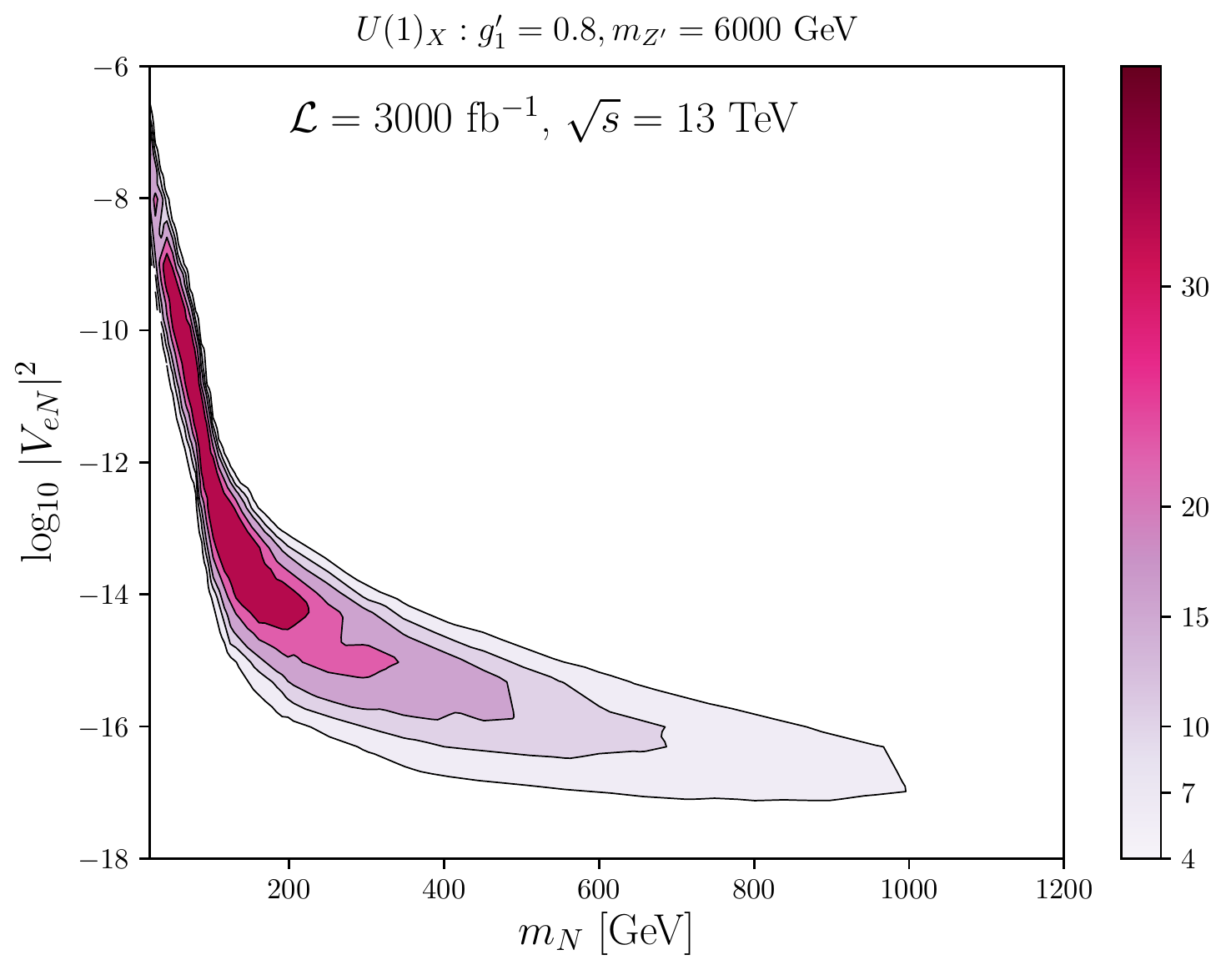}
\includegraphics[width=0.45\textwidth,angle=0]{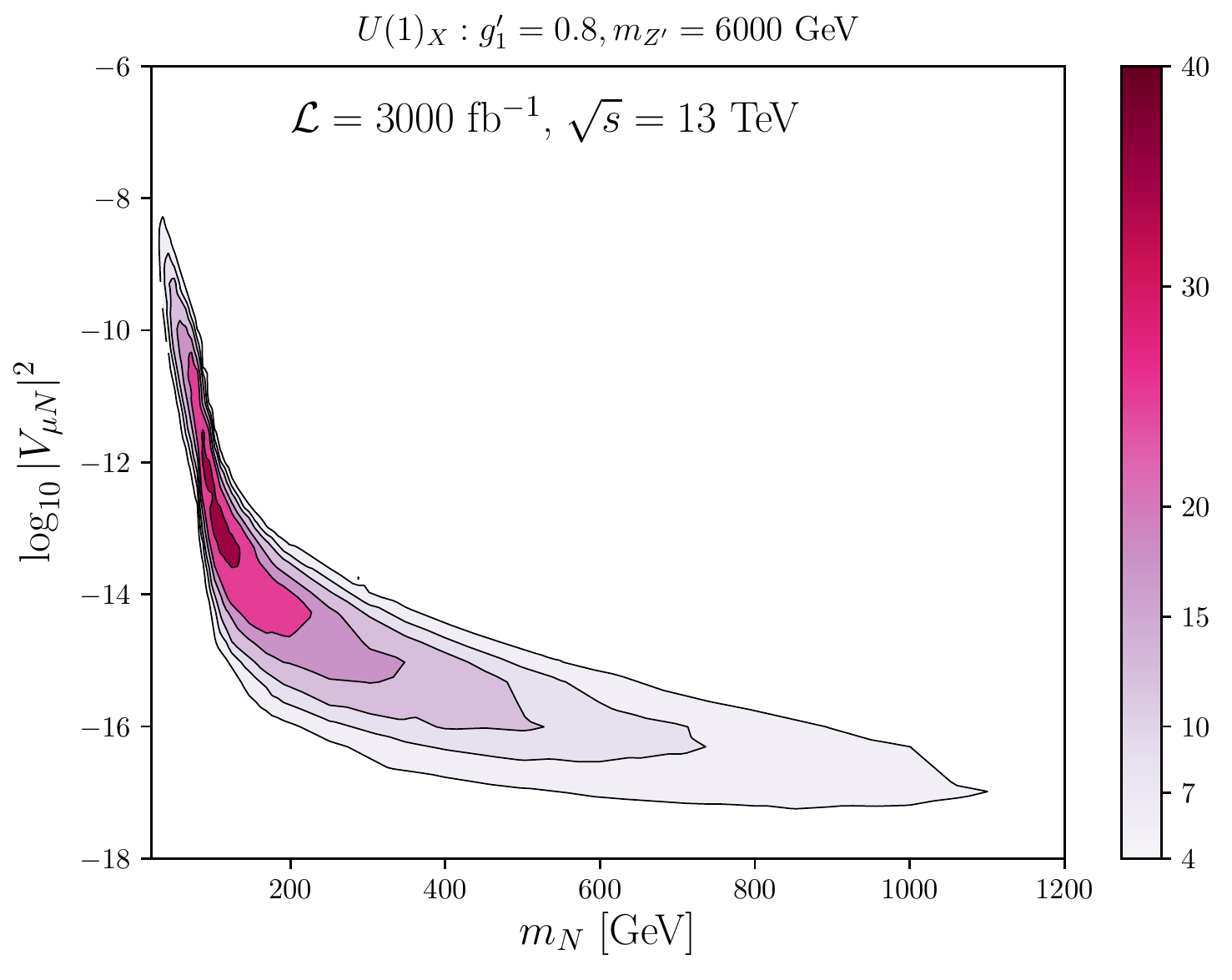}
\caption{Number of signal events for the $U(1)_{X}$ model at $\sqrt{s}=13$ TeV expected in $\mathcal{L}=3000$ fb$^{-1}$ with the ATLAS 1DV ID search. Reach for mixings in the electron (left) and muon (right) sector are shown.}
\label{nEventsATLAS1DVU1X}
\end{figure}

We show in Figure~\ref{fig:nEventsATLAS1DVBL} the estimated number of signal events at $13$ TeV and $3000$ fb$^{-1}$ for both electron and muon mixing, for the $U(1)_{B-L}$ model. One can set a $95\%$CL exclusion region with at least $3$ signal events, which is reasonable to set as a requirement for discovery in the absence of background. Analogous plots for the $U(1)_{X}$ model are shown in Figure~\ref{nEventsATLAS1DVU1X}. A larger parameter region can be excluded in the $U(1)_{X}$ model due to its higher cross section, although the strategy is sensitive to both models.  Mixings as low as $\sim 10^{-16}$ can be accessed for $m_{N}\sim 500$~GeV in the $U(1)_{B-L}$ model, and $\sim 10^{-17}$ for $m_{N}\sim1$~TeV in the $U(1)_{X}$ model.

\subsection{Reach with CMS 2DV + jets}

The above search required at least one DV reconstructed in the inner tracker. Now we explore the reach with a search requiring exactly two DVs. This has the immediate advantage of being a search free from backgrounds. In addition, with the identification of two DVs we can access the mass of the $Z'$ boson, when combining the kinematics with DV information~\cite{Cottin:2018hyf}.

The CMS search in Ref.~\cite{Sirunyan:2018pwn} looked for two DVs in the CMS inner tracker in addition to jets in the final state. The original search targeted at supersymmetric models, but the collaboration provided a reinterpretation method for extending the results to other models with pair-produced long-lived particles. Here we validate this method\footnote{Another reinterpretation of this procedure in the context of Twin Higgs Doublet models was done recently in Ref.~\cite{Alipour-fard:2018mre}.}, which is based on generator-level selections that approximately replicate the vertex-reconstruction efficiency.

We reconstruct jets with \textsc{FastJet}~3.1.3~\cite{Cacciari:2011ma} using the anti-$k_{t}$ clustering algorithm with jet radius parameter $R=0.4$. At least four jets are required with $p_{T}>20$~GeV and $|\eta|<2.5$. The variable $H_{T}$, which is the scalar sum of the $p_{T}$ of all generated jets with $p_{T}>40$~GeV, is computed and events must have $H_{T}>1000$~GeV. 

Two DVs are required, both within a transverse distance between $0.1$ and $20$~mm (near the CMS beampipe). The two DVs must also be separated from each other in the transverse plane, with distance $d_{{VV}}>0.4$~mm. All daughter particles coming from the DVs (namely, $u,d,s,c,b$ quarks and electrons, muons and tau leptons) must satisfy $p_{T}>20$~GeV and $|\eta|<2.5$, and have impact parameter $|d_{0}|\geq 0.1$~mm. In addition, the sum of the $p_{T}$ of the daughter particles has to be at least 350~GeV.  When calculating this value, we multiply the $p_{T}$ of $b$ quarks by 0.65 to account for lower reconstruction efficiency due to the long lifetime of $B$ hadrons, as instructed in Ref.~\cite{Sirunyan:2018pwn}. All selections are summarized in Table~\ref{tab:selectionsCMS}.

\begin{table}[h]
\begin{center}
\begin{tabular}{p{3.0cm}|p{11.5cm}}
\hline
Trigger         &   $H_{T}> 1000$ GeV\\
Jet selection   &  At least 4 jets with $p_{T}>20$ GeV and $|\eta|<2.5$\\

DV region       & 2 DVs within $0.1$ mm $<r_{ {DV}}<20$ mm and $d_{VV}>0.4$ mm\\
DV selection    & Made from tracks with  $|d_{0}|\geq 0.1$ mm, $p_{T}>20$ GeV and $|\eta|<2.5$. $\sum p_{T} \geq 350$ GeV, correcting for $b$ quarks. \\
\hline
\end{tabular}
\end{center}
\caption{\label{tab:selectionsCMS} Cuts for the CMS 2DV + jets search following the reinterpretation procedure in~\cite{Sirunyan:2018pwn}. }
\end{table}

As no prior validation has been done before, we validate this search on a minimal flavour violating model of $R$-parity violating supersymmetry (MFV RPV SUSY), same as the CMS benchmark signal model. We generate events in \textsc{Pythia8}~\cite{Sjostrand:2014zea} for pair production of long-lived neutralinos $\tilde{\chi}^{0}_{1}$ by quark-antiquark annihilation. The neutralino then decays into a top anti-quark and a virtual top squark, and the virtual top squark decays into strange and bottom anti-quarks. The model spectrum is generated with \textsc{SOFTSUSY 3.6.1}~\cite{Allanach:2001kg}, and is read as input to \textsc{Pythia8} using the SLHA structure~\cite{Allanach:2008qq}. 

\begin{figure}[h]
\centering
\includegraphics[width=0.7\textwidth,angle=0]{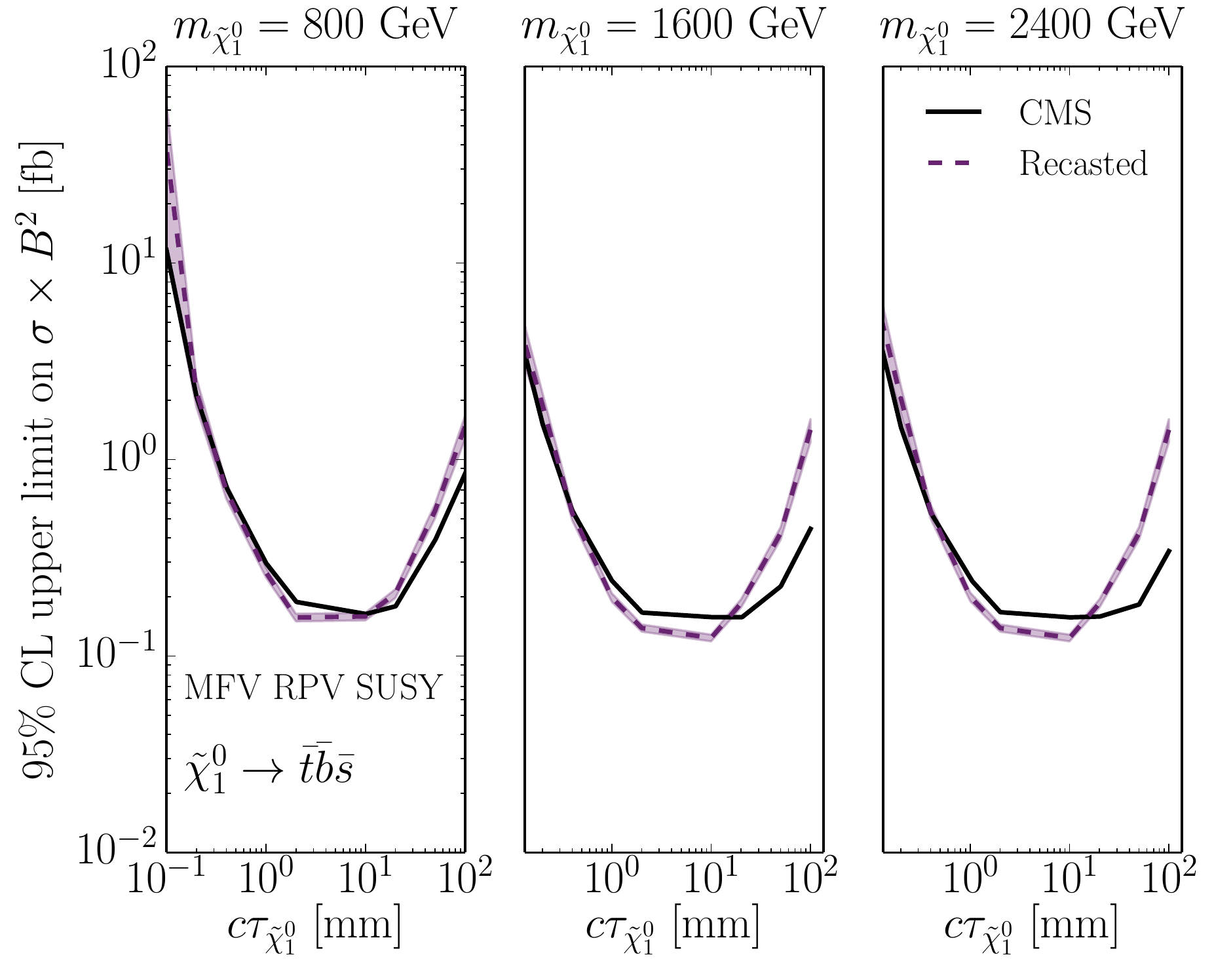}
\caption{Validation of the $95\%$ CL observed CMS upper limits~\cite{Sirunyan:2018pwn} for three mass points in the MFV RPV SUSY model: $m_{\tilde{\chi}^{0}_{1}}=800, 1600, 2400$~GeV. }
\label{fig:CMS2DVVal}
\end{figure}

Figure~\ref{fig:CMS2DVVal} shows our recast $95\%$ CL limit taken with zero background and 3 signal events, against the CMS exclusion for three different benchmarks.

\begin{figure}[h]
\centering
\includegraphics[width=0.7\textwidth,angle=0]{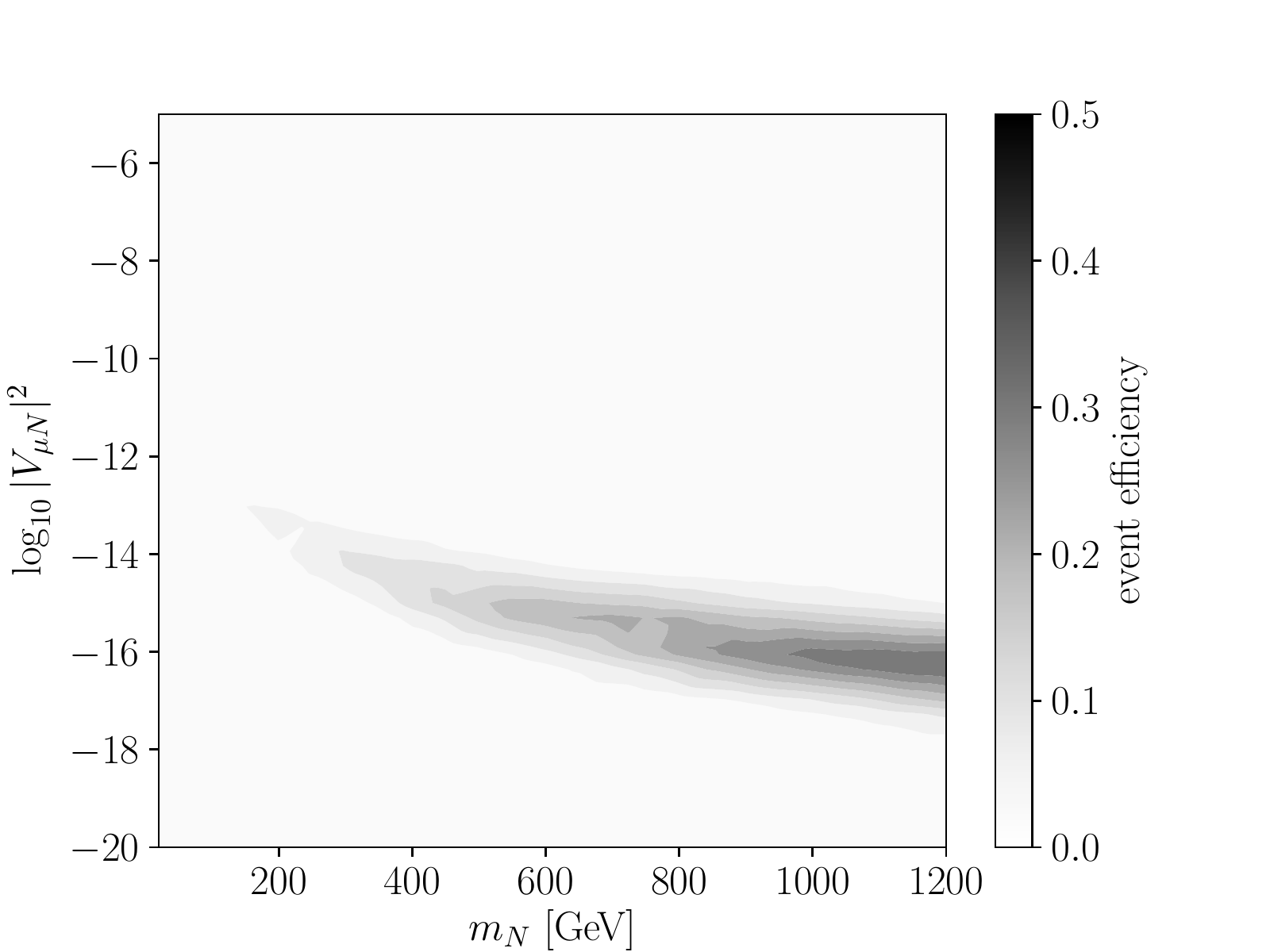}
\caption{Representative event level efficiency of the CMS 2DV + jets search as a function of $m_{N}$ and $|V_{\mu N}|^{2}$.  }
\label{fig:CMS2DVeff}
\end{figure}

The event-level efficiency is shown in Figure~\ref{fig:CMS2DVeff} for the $U(1)_{X}$ model. We note the same pattern as with the ATLAS 1DV ID search, although the sensitivity here is affected by the mass ratio between $m_{N}$ and $m_{Z'}$, as the angular opening of the decay products of the long-lived neutrino is proportional to its boost. This means that for neutrino masses roughly around 100~GeV and below, it is harder for the decay products to be effectively resolved into our reconstructed jets, failing the selection criteria. The additional condition that both neutrinos must decay well separated near the CMS beampipe imposes another strict requirement on our displaced events. 

\begin{figure}[h]
\centering
\includegraphics[width=0.45\textwidth,angle=0]{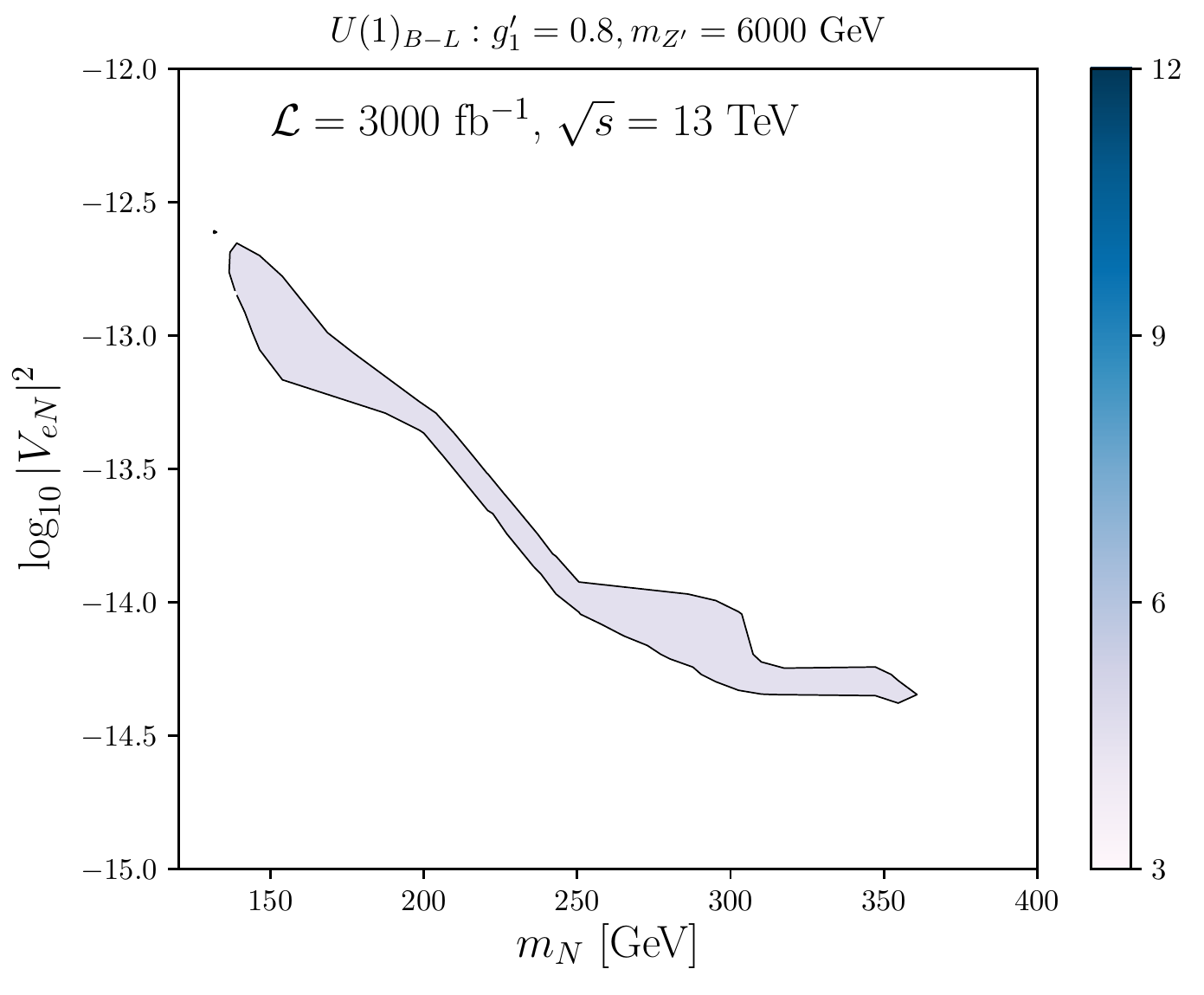}
\includegraphics[width=0.45\textwidth,angle=0]{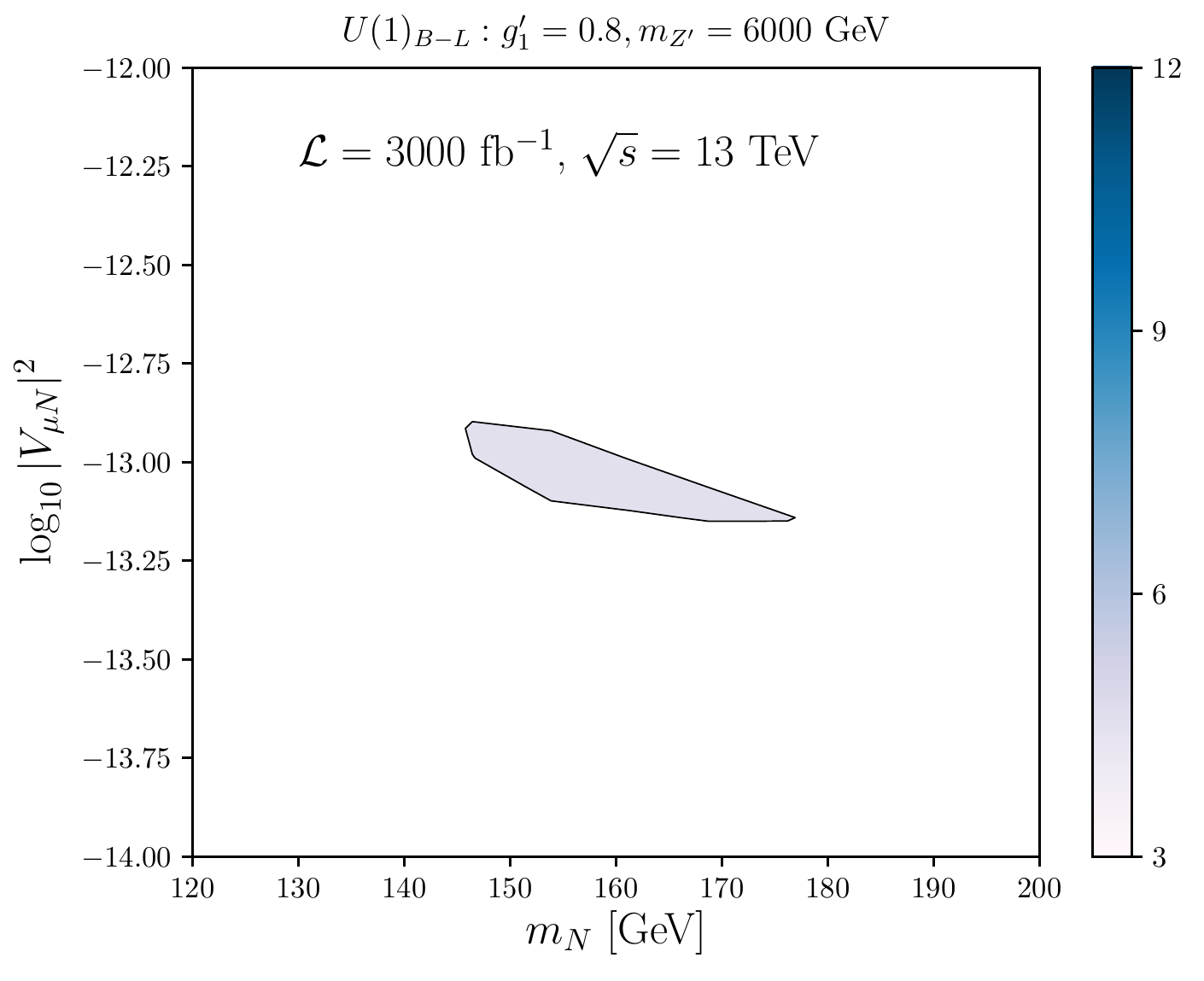}
\caption{Number of signal events for the $U(1)_{B-L}$ model at $\sqrt{s}=13$~TeV expected for $\mathcal{L}=3000$ fb$^{-1}$ with the CMS 2DV+jets strategy. Reach for mixings in the electron (left) and muon (right) sector are shown.}
\label{fig:CMSBLreach}
\end{figure}

\begin{figure}[h]
\centering
\includegraphics[width=0.45\textwidth,angle=0]{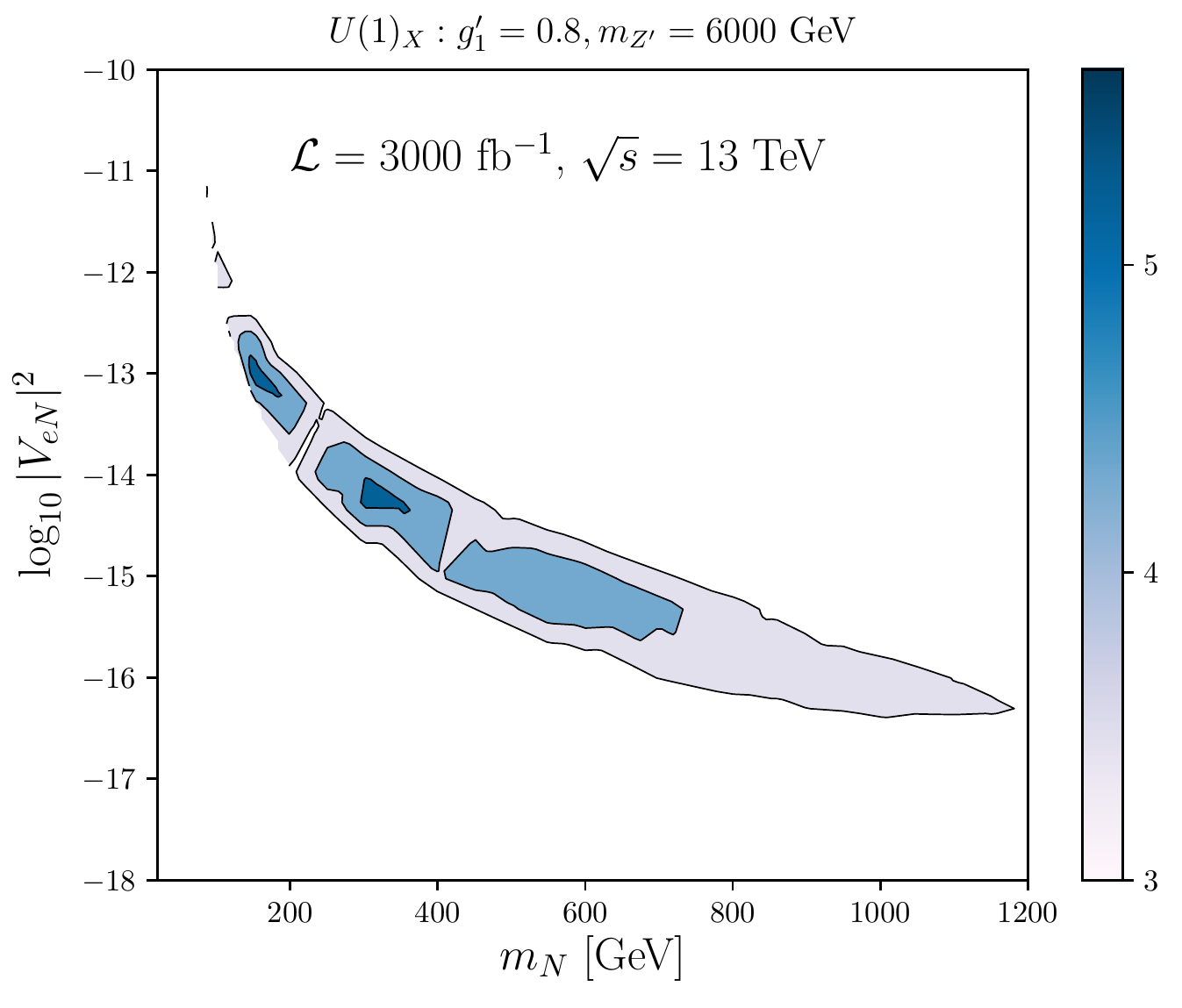}
\includegraphics[width=0.45\textwidth,angle=0]{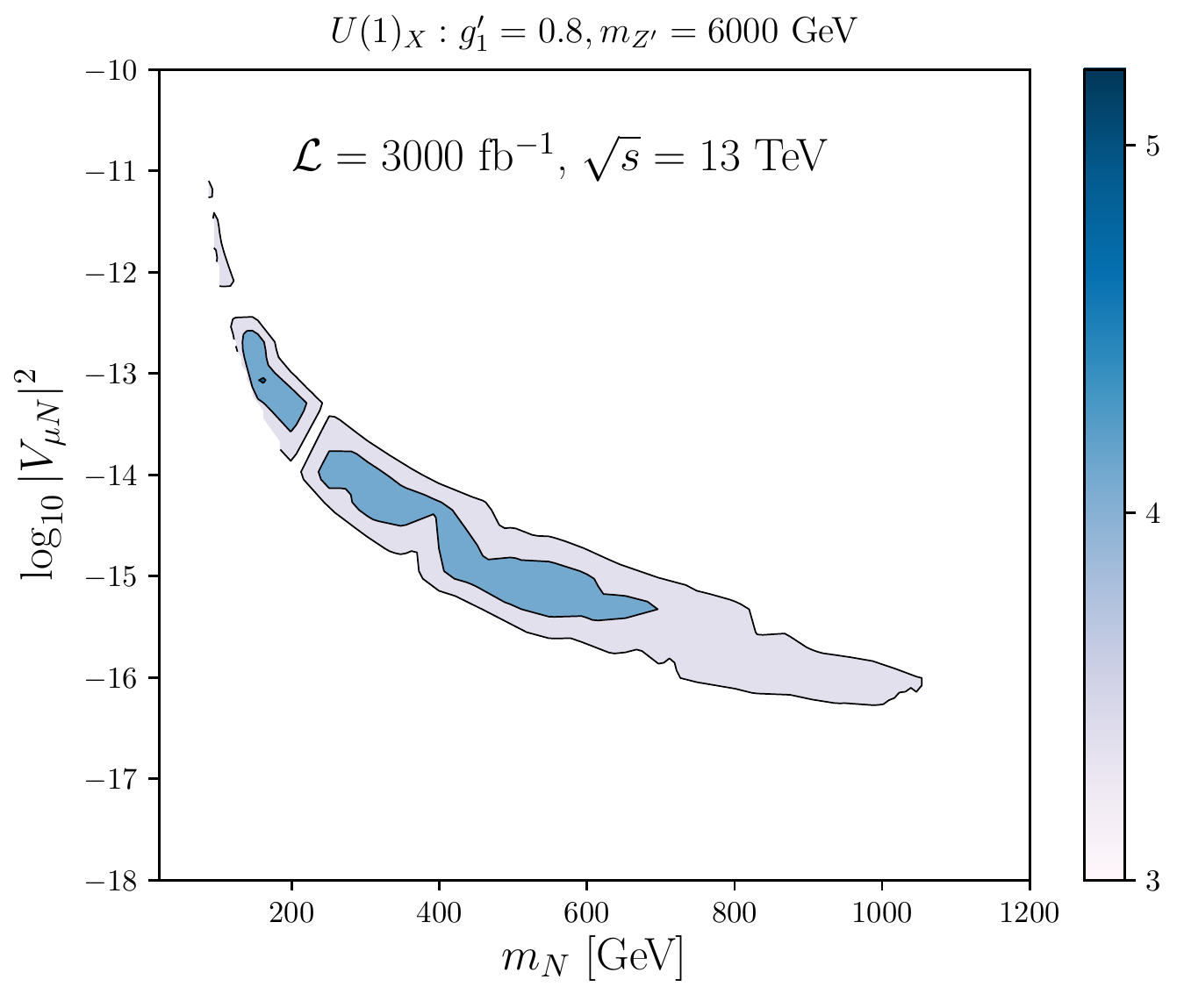}

\caption{Number of signal events for the $U(1)_{X}$ model at $\sqrt{s}=13$~TeV expected for $\mathcal{L}=3000$ fb$^{-1}$ with the CMS 2DV+jets strategy. Reach for mixings in the electron (left) and muon (right) sector are shown.}
\label{fig:CMSU1Xreach}
\end{figure}

When estimating the number of signal events, we note nearly no sensitivity to the $B-L$ scenario, as shown in Figure~\ref{fig:CMSBLreach}. The situation improves in the $U(1)_{X}$ model shown in Figure~\ref{fig:CMSU1Xreach}, although the parameter space of the models that can be accessed can also be covered by the ATLAS 1DV search, except for a small region at small mixings and TeV masses. We can reach mixings down to $\approx 10^{-17}$ for $m_{N}=1200$ GeV in the electron case. The mass reach in $m_{N}$ is limited by its proper lifetime. As for mixings above $\approx 10^{-15}$ and masses higher than $\approx 1$~TeV, the heavy neutrino is already decaying too promptly.

\clearpage
\section{Conclusions}
\label{close}
We have studied discovery prospects at the LHC of a new right-handed neutrino, which can be pair produced via a $Z'$ gauge boson. The neutrino has a macroscopic lifetime to be able to decay inside the inner trackers of the LHC detectors to leptons and jets, enabling the reconstruction of displaced vertices from its charged tracks.

We focus on two models, namely the minimal $U(1)_{B-L}$ model and its $U(1)_{X}$ extension, allowing a more general quantum charge assignments for an increased $Z'$ production cross section.
We examine discovery prospects on a benchmark scenario near the limit of current lepton resonance searches at the LHC, with the explicit parameters $(m_{Z'},g'_{1})= (6$ TeV, $0.8$). We study production and decay of the heavy displaced neutrino to leptons and jets, and reinterpret existing searches for displaced vertices. Our focus is on the current inner tracker DV searches: an ATLAS search for at least one DV (ATLAS 1DV ID) and a CMS search for exactly two DVs in multi-jet events (CMS 2DV+jets). We find that the ATLAS 1DV ID is the most sensitive across a wide range in the light-heavy neutrino mixing and mass space. Mixings as low as $\approx 10^{-17}$ can be accessible for a heavy neutrino of mass around hundreds of GeV.

By considering displaced vertices reconstructed inside the muon spectrometer (MS)~\cite{Aaboud:2017iio}, one should be able to further constrain the parameter space of these heavy neutrino models. The scenario with heavy neutrino masses lower than the ones considered in this work was studied first in Ref.~\cite{Batell:2016zod} and later in Ref.~\cite{Deppisch:2018eth}, the latter focusing on neutrinos which were pair produced from decays of the Higgs boson. The authors show that mixings with muons as small as $\approx 10^{-14}$ can be probed at the LHC for heavy neutrino masses below 60~GeV. In our case, the MS searches would be sensitive to even smaller values of the mixing, below $\approx 10^{-20}$ (see Figure~\ref{fig:ctau}). In view of the challenges in simulating vertex reconstruction in the muon spectrometer, estimation of displaced efficiencies and proper treatment of backgrounds, we leave this study for a future work.

\section*{Acknowledgments}
G.C. would like to thank the Korea Institute of Advance Studies (KIAS) for hospitality offered (at the time of ICHEP 2018) when this work was initiated, Brian Shuve for useful comments and Gabriel Torrealba for technical advice with python. The research of C.-W. C. was supported in part by the Ministry of Science and Technology (MOST) of Taiwan under Grant No. MOST-104-2628-M-002-014-MY4 and MOST-108-2112-M-002-005-MY3. G.C. acknowledges support by Grant No. MOST-107-2811-M-002-3120 and CONICYT-Chile FONDECYT Grant No. 3190051. The work of A. D. is supported by the Japan Society for the Promotion of Science (JSPS) Post- doctoral Fellowship for Research in Japan.
\section*{Appendix} 
We write the partial decay widths of the $Z^\prime$ as a function of $x_H$ and $g'_1$ taking $x_\Phi=1$. The partial decay width of the $Z^\prime$ into a pair of light neutrinos:
\bea
\Gamma(Z' \to 2 \nu) = \frac{m_{Z'}}{24\pi}~g_L^\nu \Big[g'_1, x_H \Big]^2,
\eea
with $g_L^\nu[g'_1, x_H] \equiv \Big((-\frac{1}{2}) x_H + (-1)\Big) g'_1$. The partial decay width of the $Z^\prime$ into a pair of charged leptons: 
\bea
\Gamma(Z' \to 2 \ell) = \frac{m_{Z'}}{24\pi}~\Big(g_L^e \Big[g'_1, x_H \Big]^2 + g_R^e \Big[g'_1, x_H \Big]^2\Big),
\eea
with $g_L^e =\Big((-\frac{1}{2}) x_H + (-1)\Big) g'_1$ and $g_R^e =\Big((-1) x_H + (-1)\Big) g'_1$.  The partial decay width of the $Z^\prime$ into a pair of up-type quarks: 
\bea
\Gamma(Z' \to 2 u) = \frac{m_{Z'}}{24\pi}~\Big(g_L^u \Big[g'_1, x_H \Big]^2 + g_R^u \Big[g'_1, x_H \Big]^2\Big),
\eea
with $g_L^u =\Big((\frac{1}{6}) x_H + (\frac{1}{3})\Big) g'_1$ and $g_R^u =\Big((\frac{2}{3}) x_H + (\frac{1}{3})\Big) g'_1$. 
The partial decay width of the $Z^\prime$ into a pair of down-type quarks: 
\bea
\Gamma(Z' \to 2 d) = \frac{m_{Z'}}{24\pi}~\Big(g_L^d \Big[g'_1, x_H \Big]^2 + g_R^d \Big[g'_1, x_H \Big]^2\Big),
\eea
with $g_L^d =\Big((\frac{1}{6}) x_H + (\frac{1}{3})\Big) g'_1$ and $g_R^d =\Big((-\frac{1}{3}) x_H + (\frac{1}{3})\Big) g'_1$. In these partial decay widths, terms involving the fermion masses are negligible because they are suppressed by $m_{Z'}$, which we take to be $6$~TeV in this analysis. The partial decay width of the $Z'$ into a pair of Majorana heavy neutrinos for $m_{Z'} > 2 m_N$ (at least) is 
\bea
\Gamma(Z' \to 2 N) = \frac{m_{Z'}}{24\pi}~g_R^N \Big[g'_1, x_H \Big]^2 \Big(1- 4 \frac{m_N^2}{m_{Z'}^2}\Big)^{\frac{3}{2}},
\eea
with $g_R^N [g'_1, x_H] = \Big((0) x_H+( -1)\Big) g'_1$. The partial decay widths of the heavy neutrino can be found in Ref.~\cite{Das:2018tbd} for different masses, lighter and heavier than the SM gauge bosons.
\bibliographystyle{JHEP}
\bibliography{main}
\end{document}